\documentclass[amsmath,amssymb,amsthm,superscriptaddress,showpacs]{elsarticle}
\usepackage{graphicx}
\usepackage{lineno}

\begin{document}

%\linenumbers

\begin{frontmatter}

\title{Manageable to unmanageable transition in a fractal model of project networks}
\author{Alexei Vazquez\corref{cor}} 
\ead{alexei@nodeslinks.com}
%\cortext[cor]{}
\affiliation{Nodes \& Links Ltd, Salisbury House, Station Road, Cambridge, CB1 2LA, UK}

\begin{abstract}
Project networks are characterized by power law degree distributions, a property that is known to promote spreading. In contrast, the longest path length of project networks scales algebraically with the network size, which improves the impact of random interventions. Using the duplication-split model of project networks, I provide convincing evidence that project networks are fractal networks. The average distance between nodes scales as $\langle d\rangle \sim N^{\beta}$ with $0<\beta<1$. The average number of nodes $\langle N\rangle_d$ within a distance $d$ scales as $\langle N\rangle_d\sim d^{D_f}$, with a fractal dimension $D_f=1/\beta>1$. Furthermore, I demonstrate that the duplication-split networks are fragile for duplication rates $q<q_c=1/2$: The size of the giant out-component decreases with increasing the network size for any site occupancy probability less than 1. In contrast, they exhibit a non trivial percolation threshold $0<p_c<1$ for $q>q_c$, in spite the mean out-degree diverges with increasing the network size. I conclude the project networks generated by the duplication-split model are manageable for $q<q_c$ and unmanageable otherwise.   
\end{abstract}

\end{frontmatter}

\section{Introduction}

A project is a collection of interdependent activities to achieve a goal in a pre-determined amount of time. However, very often projects do not proceed as planned. Delays in some activities propagate to downstream activities ultimately causing a delay in the project delivery and increased costs \cite{majerowicz16}. To mitigate the damages caused by project delays we must first understand the nature of delay propagation along project networks. It has been suggested that project delays propagate along project networks following similar features to other spreading phenomena. That the mitigation of delay cascades cascades has many similarities with interventions to halt the spreading of infectious diseases \cite{ellinas19}. In turn, it is well stablished that spreading on a network can be mapped to a percolation problem \cite{newman02a}.

From our current understanding of spreading phenomena and percolation on random networks we can envision projects where minimal intervention strategies would be sufficient to halt delay cascades. I will call such projects manageable projects. In contrast, there can be projects where stopping the propagation of delay may require interventions at a finite percent of activities. I will call the latter unmanageable project networks. Our scope is then to identify what project networks are manageable or unmanageable, using percolation theory.

Both the shape of the degree distribution and the scaling of the network diameter $d_{\max}$ with the network size $N$ are relevant to understand the percolation properties of networks. In percolation we assume nodes are present with some probability $p$, otherwise absent due to failure or intervention \cite{stauffer85}. For a chain of nodes we have $d_{\max}\sim N$ and for any $p<1$ the chain breaks it into finite segments. In that case we say the percolation threshold is $p_c=1$. We require all nodes to be present to obtain a connected subnetwork with a size proportional to the whole network. In contrast, in geometrical lattices embedded in a space of dimension $D>1$, such as the Manhattan grid in $D=2$, there is a percolation threshold $0<p_c<1$ \cite{stauffer85}.

In the early 2000s it was discovered that undirected random networks with a power law degree distribution $P(k)\sim k^{-\gamma}$ do not have a percolation threshold when $\gamma<3$ \cite{albert00,callaway00}. For $\gamma<3$ the effective reproductive number $\langle k(k-1)\rangle / \langle k\rangle$ diverges with increasing the network size. As a consequence, for any node occupation probability $p>p_c=0$ the network has a giant component, reflecting a robustness against random failure. Similar results where obtained for spreading dynamics on random networks \cite{pastor-satorras01}, where the epidemic threshold plays the role of the critical node occupancy.

The absence of an epidemic threshold was challenged by the deactivation model \cite{klemm02}, which can generate networks with $\gamma<3$ but they have a finite epidemic threshold. We later showed that the deactivation model has a one-dimensional topology \cite{vazquez03}. From these studies we learned that the divergence of the reproductive number is not a sufficient condition for the absence of percolation threshold. It was argued that we require the small-world property as well: a logarithmic scaling of the network diameter with the network size ($d_{\max}\sim\log N$) \cite{boguna03,vazquez03b}. Finally, the necessity of the small world property can be generalized to properties associated with the hierarchy of nested subgraphs \cite{serrano11}.

Fractals are somewhere in between networks embedded in some integer dimension space and random networks. In fractals the length scales with the size as $L\sim N^{1/D_f}$, where $D_f$ is the fractal dimension and takes non-integer values.  Deterministic fractals like the Sierpinski gasket ($D_f\approx 1.6$) and the Koch curve ($D_f\approx 1.3$) are fragile, with a site and node percolation threshold $p_c=1$ \cite{gefen84a,taitelbaum90}. In the case of recursive and hierarchical networks with power law degree distributions the percolation properties depend on the scaling between the network diameter and the network size, as observed for random networks \cite{rozenfeld07}. When there is a fractal scaling ($d_{\max}\sim N^\beta$) the studied networks have a finite percolation threshold $0<p_c<1$ as it is the case for regular lattices. In contrast, when they exhibit the small-world scaling ($d_{\max}\sim\log N$) they have $p_c=0$ as in random scale-free networks. For a comprehensive review on the topic see Refs. \cite{dorogovtsev08,cruz23}.

Recently we have reported that projects may be a new class of networks \cite{vazquez23}. The network nodes are discrete activities carried on by a single individual or team.  The network links are logical pairwise relations between activities. The directed link (arc) A$\rightarrow$B indicates that activity A most be finished before activity B can start. The activities require some duration to complete their execution. The activities durations and relations are used to build the project timeline or plan. The analysis of construction project networks reveals two fundamental properties: (i) identical scale-free distributions of out- and in-degrees and (ii) an algebraic scaling of the longest path size with the network size $L\sim N^{1-0.79}$ \cite{vazquez23}. We also introduced a network growth model based on node duplication and split that explains both properties \cite{vazquez23}.

Here I demonstrate that networks generated by the duplication-split model are fractals with non-trivial percolation properties. The work is organized as follows. In Sec. \ref{model} we introduce the duplication-split model. Next we analyze the number of nodes vs distance scaling in Sec \ref{md} and the percolation properties in Sec. \ref{percolation}. The final conclusions are reported in Sec. \ref{conclusions}.

\begin{figure}[t]
\includegraphics[width=4.75in]{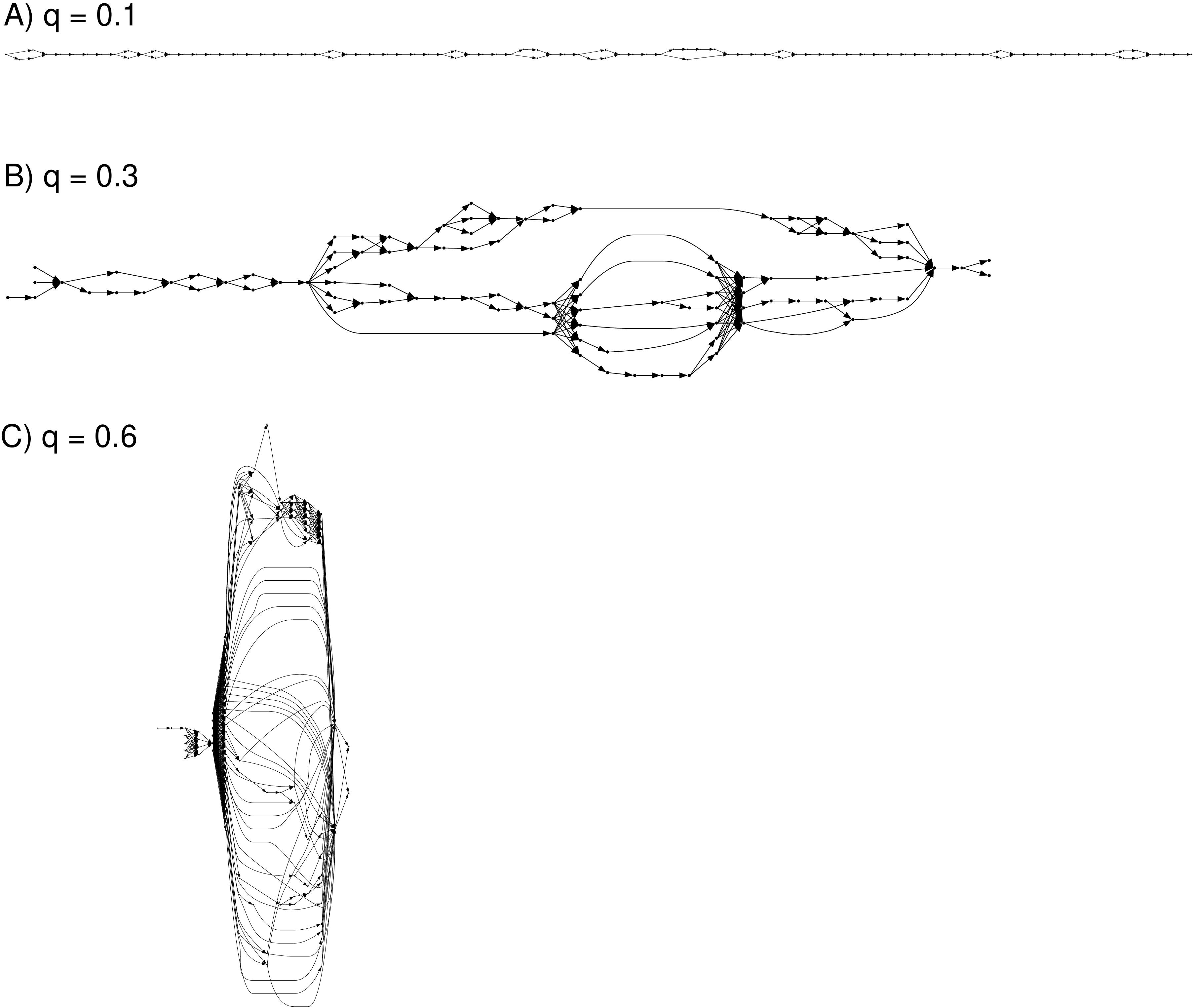}
\caption{Activity networks generated by the duplication-split model with 100 nodes and A) $q=0.1$, B) $q=0.3$ and C) $q=0.6$. The nodes layout was generated with graphviz.}
\label{fig_model}
\end{figure}

\section{Model definition}
\label{model}

The duplication-split model is defined as follows. Initial condition: We start with a given activity network, which can be of small size. I use the minimal network A$\rightarrow$B. Evolution: At each step, we select a node at random and with probability $q$ we duplicate the node, otherwise we split the node. The duplication rule creates a new copy of the selected node, inheriting all its incoming and outgoing arcs. The split rule creates two nodes out of the selected node, where one inherits the incoming arcs and the other the outgoing arcs of the original node, plus an additional arc between the new nodes. Duplication models specialization into independent (duplication) or dependent (split) activities \cite{vazquez23}. Figure \ref{fig_model} shows
duplication-split networks with 100 nodes and various duplication rates. The network for $q=0.1$ is close to a linear chain. The network for $q=0.3$ is a mix of chains and parallel paths. The network for $q=0.6$ is characterized by several parallel paths and a short distance from the sources at the left to the sinks at the right. These networks are directed and acyclic. Finally, the nodes out-degree (number of predecessors) and in-degree (number of successors) follow the same power law distribution $P(k)\sim k^{-1/q}$ \cite{vazquez23}. 

\begin{figure}
\begin{center}
\includegraphics[width=3.3in]{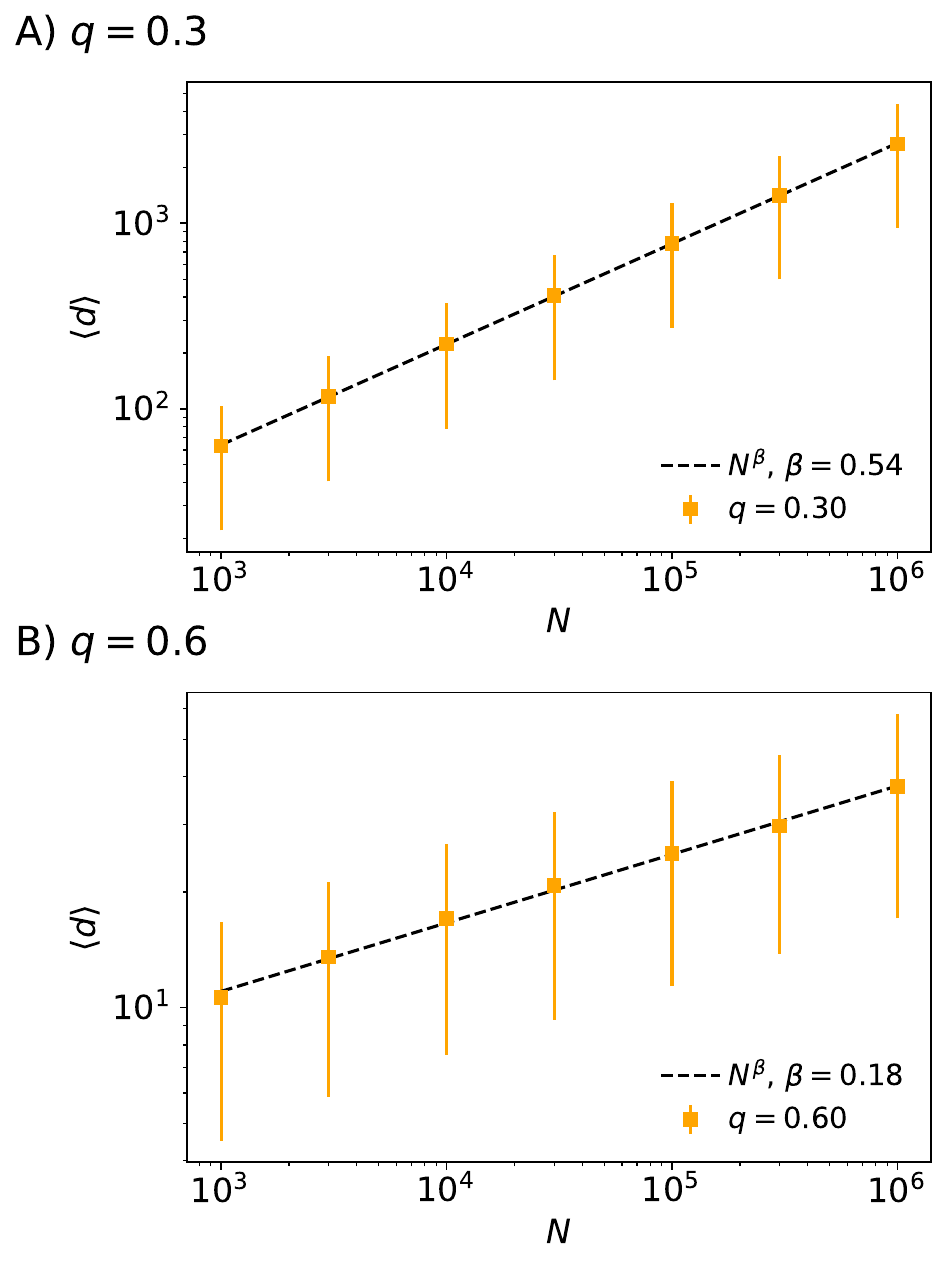}
\end{center}
\caption{Scaling of the average distance $\langle d\rangle$ from root nodes to reachable nodes with the network size $N$. The symbols represent an average over all roots in the network and 100 network realizations. The lines are a fit to a power law scalings. The error bars represent the standard deviation, including variations between nodes in the same network and across network realizations.}
\label{fig_d_n}
\end{figure}

\section{Number of nodes scaling with distance}
\label{md}

First we investigate the scaling of the average distance between the network roots and any other node. The roots of directed acyclic graphs are the nodes with in-degree equal 0. Duplication-split networks can have multiple roots, as shown in the Fig. \ref{fig_model}B. Starting from a root we calculate the number of nodes at a distance $d$. We repeat the process for every root and take the average. Finally, we average over multiple network realizations. From that distribution we calculate the average distance $\langle d\rangle$ from the root to the reachable nodes. Figure \ref{fig_d_n} reports the scaling of $\langle d\rangle$ vs the network size $N$. There is an evident power law scaling $\langle d\rangle \sim N^\beta$ with $0<\beta<1$. This result is in agreement with our previous analysis for the length of the longest path \cite{vazquez23}. 

Next we focus on the scaling of the average number of nodes within a distance $d$ to a root $\langle N\rangle_d$.
When the distances scale as $\langle d\rangle \sim N^\beta$ we expect the average number of nodes to scale as $\langle N\rangle_d \sim d^{D_f}$ with $D_f=1/\beta$. If a network is a $D$ dimensional object, then  $\langle N_d\rangle \sim d^D$, where $D$ is an integer larger than 0. In contrast, if $\langle N\rangle_d\sim d^{D_f}$, where $D_f$ is a non-integer number, the network is a fractal with a fractal dimension $D_f$.

\begin{figure}
\begin{center}
\includegraphics[width=3.3in]{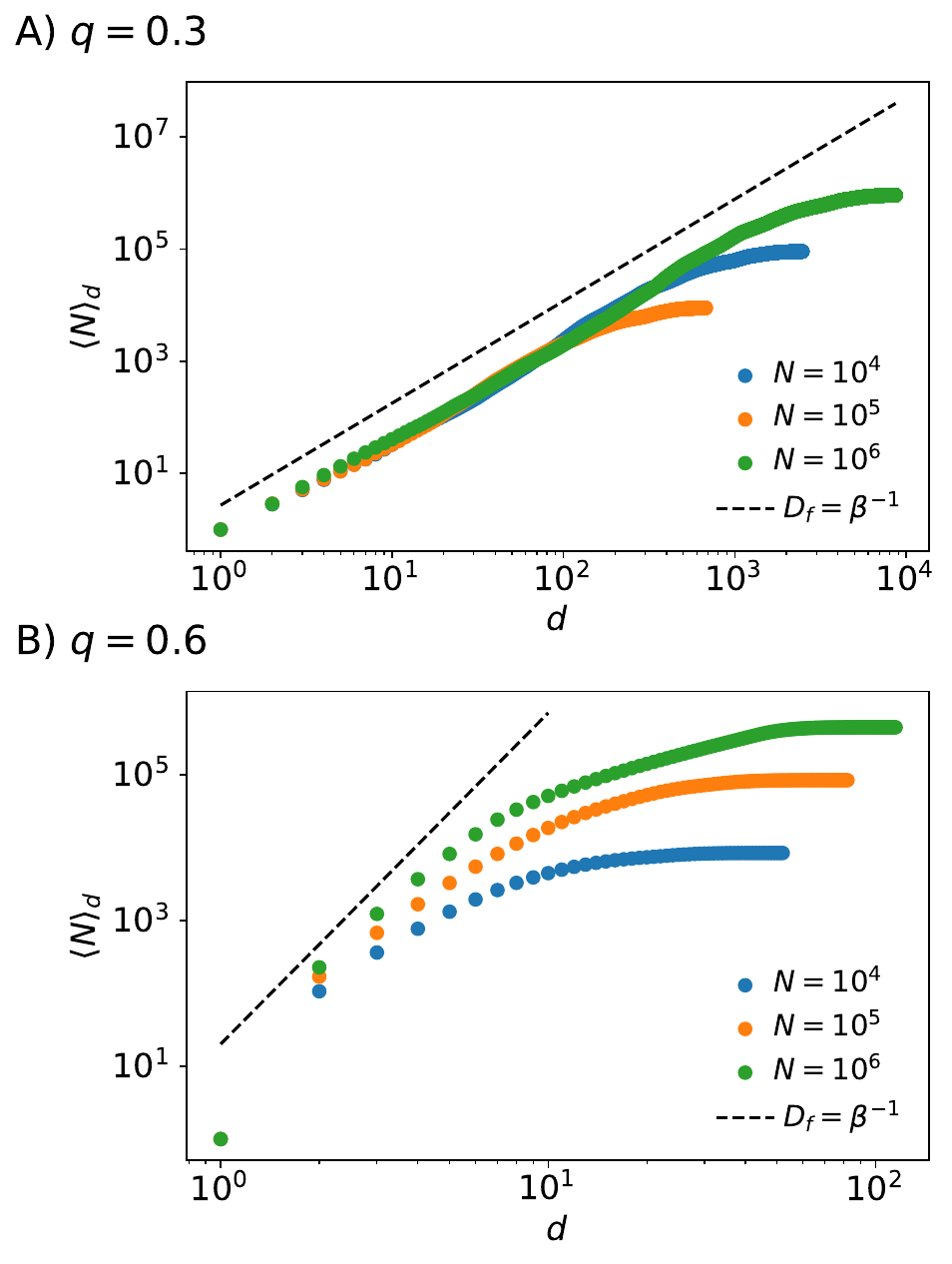}
\end{center}
\caption{Average number of nodes scaling with the distance in duplication-split networks. A) $q=0.3$ and average over 10 networks. B) $q=0.6$ and one network.}
\label{fig_m_d}
\end{figure}

Figure \ref{fig_m_d}A reports $\langle N\rangle_d$ vs $d$ scaling for duplication-split networks with $q=0.3$. There is an evident linear relationship in the log-log scale with a slope $D_f=1/\beta\approx 1.8$, where $\beta$ was obtained from the linear fit in Fig. \ref{fig_d_n}A. When we increase the network size from $10^4$ to $10^6$ we observe an extension of the region with linear scaling. The deviation at large $d$ is a finite size effect. The estimated fractal dimension is larger than 1 and it is non-integer, indicating that duplication split models are fractals. For $q=0.6$ the plot $\langle N\rangle_d$ vs $d$ exhibits strong finite size effects (Fig. \ref{fig_m_d}B). As the network size increases, the numerical points of $\langle N\rangle_d$ vs $d$ converges to the expected $\langle N\rangle_d\sim d^{D_f}$ scaling with $D_f=1/\beta$.  

\begin{figure}[t]
\begin{center}
\includegraphics[width=3.3in]{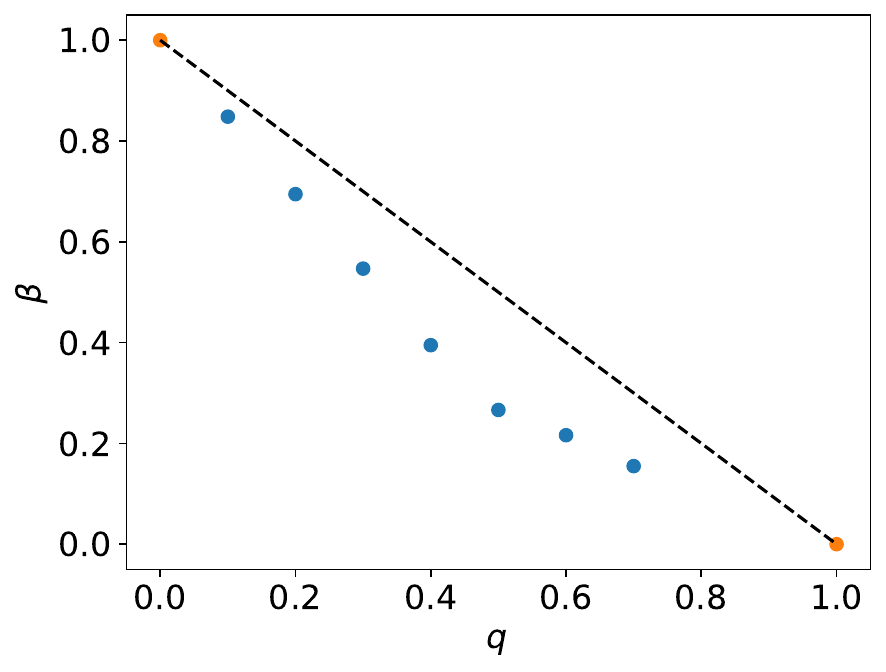}
\end{center}
\caption{The value of $\beta$ for different values of $q$, as estimated from the slope of $\log \langle d\rangle$ vs $\log N$. The two extreme points at $q=0$ and $1$ are the expectations for a linear chain and a network where the distance does not change with increasing the network size, respectively. For $q=0.8$ and 0.9 we could not finish the analysis for the network size $N=10^6$. This large or larger networks sizes are required due to the very slow increase of $\langle d\rangle$ with $N$ when $q$ approaches 1.}
\label{fig_alpha_q}
\end{figure}

We repeat this calculation for duplication-split models with different values of $q$. For $q\leq0.6$ the networks were grown to $N=10^3$, $10^4$, $10^5$ and $10^6$ and we took 10 network realizations for each value of $q$. For $q>0.6$ and $N=10^6$ the simulations take extremely long times, so we took one network realization.  The scaling between number of nodes and distances is more evident from the $\langle d\rangle$ vs $N$ plot (Fig. \ref{fig_d_n}). Therefore we used this plot to estimate $\beta$. The fractal dimension can be deduced from the equation  $D_f=1/\beta$. Figure \ref{fig_alpha_q} reports the $\beta$ estimates as a function of $q$. As the duplication rate increases $\beta$ decreases, indicating a slower power law dependency of $\langle d\rangle$ vs $N$. As a reference we show the line $1-q$, representing the tendency of a path to grow due to the split rule. The length of a path grows as $dl/dN = (1-q)l/N$, resulting in the scaling $l\sim N^{1-q}$ \cite{vazquez23}. However, the shortest path has a slower dependency with $N$, because there are multiple paths between two nodes. 

\section{Percolation}
\label{percolation}

The propagation of risk in project networks is a major concern to project management. For example, an activity may be delayed and, as a consequence, shift the finish dates of downstream activities. In turn, project managers deploy intervention strategies to halt the risk propagation in the activity network. This phenomenology can be mapped to the mathematical problem of percolation in the activity network.

The mapping proceed as follows. Suppose the project manager deploys intervention strategies at a fraction $f$ of the nodes. These are the nodes that will fail to transmit delay, in a similar manner to the concept of immunization in the context of disease spreading. In contrast, the remaining fraction $p=1-f$ of nodes will transmit delays. By removing the nodes that do not transmit delay we fragment the network into clusters of nodes. That means a delay in any node on a given cluster will only propagate to other nodes within the cluster. If there is a cluster with a size proportional to the network size, a giant component, then we haven't mitigated the possibility of delay cascades. Otherwise, if all clusters have a finite size that does not scale with the network size, then we have contained the delay propagation to clusters of finite size. Therefore, by investigating the percolation properties as a function of the site occupation probability $p$ and the network size $N$, we gain an understanding about the effort required to mitigate the delay cascades.

Furthermore, I define a network as {\em manageable} if it takes interventions at a vanishing fraction of nodes $f$ to mitigate delay cascades, where vanishing  means $f\rightarrow 0$ when $N\rightarrow\infty$. In percolation that is equivalent to say that we need a site occupation probability $p_c=1$ to obtain a giant component. Removal of a finite number of nodes breaks the network into clusters of finite size. In contrast, I define a network {\em unmanageable} if it takes interventions at a non-vanishing fraction of nodes to mitigate delay cascades,  where non-vanishing  means $f\rightarrow {\rm const.}$ when $N\rightarrow\infty$. In percolation that is equivalent to say there is a site occupation probability $0<p_c<1$ to obtain a giant component. It takes the removal of a finite fraction of nodes to break the network into clusters of finite size. The cost of mitigation is of the order of the network size and therefore large.

In the following we investigate the percolation properties of the duplication-split networks. The duplication-split networks are directed. In directed networks we can talk about out-components, in-components and strongly connected components. Given a node, its out-component (in-component) is the sub-network of nodes reachable following the directed links in their forward (backward) direction. When the in- and out-component coincide then we have a strongly connected component. Duplication-split networks are acyclic and therefore they do not contain strongly connected components.

In duplication split networks the roots, the nodes with in-degree zero, have giant out-components with a size proportional to the network size. That implies that risk can propagate along the giant component affecting a number of nodes comparable to the network size. Now, let's say a fraction $1-p$ of the nodes do not propagate the risk due to interventions, while a fraction $p$ transmits risk. That is, out-component site percolation.

For random directed networks, the site percolation thresholds is $p_c = 1/\langle k_{out}\rangle$, where $\langle k_{out}\rangle$ denotes the average out-degree \cite{schwartz02}. The duplication-split networks have the out-degree distribution $P(k_{out})\sim k_{out} ^{-1/q}$ \cite{vazquez23}, with a finite  or diverging average out-degree for $q<q_c=1/2$ or $q>q_c$, respectively. If duplication-split networks would behave like random directed networks, we  would expect a finite percolation threshold $0<p_c<1$ for $q<q_c$ and for $q>q_c$ no percolation threshold ($p_c=0$). We will address these expectations by means of numerical simulations.

\begin{figure}[t]
\includegraphics[width=4.75in]{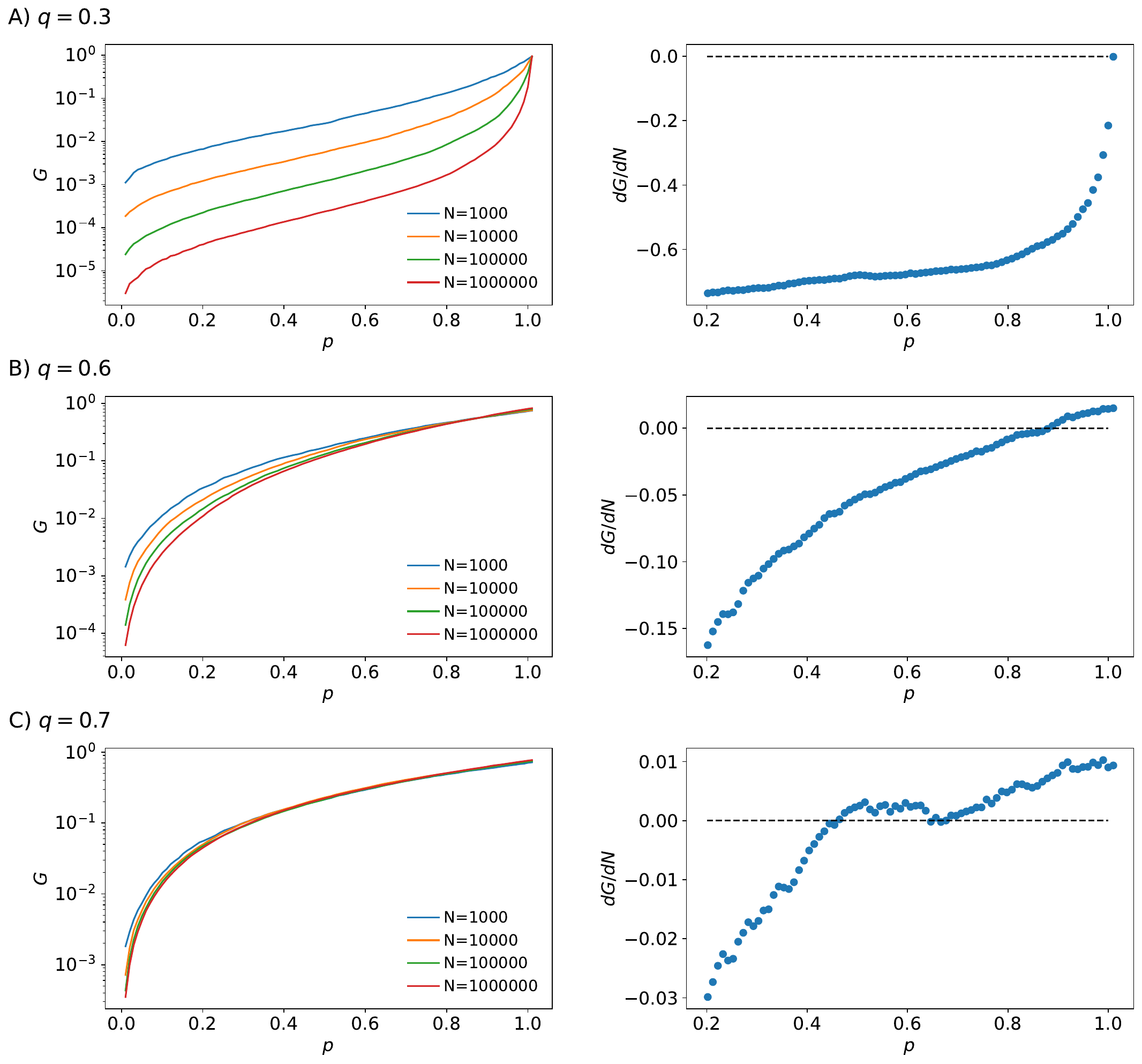}
\caption{Fraction of nodes in the giant out-component $G$ and its size-dependent slope $dG/dN$ as a function of the node occupancy $p$, for duplication-split networks with different duplication rates $q$ and network size $N$.  All data points represent averages over 10 networks and 10 sets of node occupancies.}
\label{fig_p_q}
\end{figure}

We generated duplication split-networks and calculated their giant component for node occupation probabilities $0<p\leq1$. For each value of $p$ we averaged over 10 networks and 10 realizations of nodes occupations. The duplication-split networks are directed and acyclic and therefore we focus on the out-component sizes. We generate the duplication-split networks and remove each node with probability $1-p$. Then we loop over the nodes in their topological order. If the node has not been visited, we do a breadth-first search of all reachable nodes following the forward direction of the directed links and record this component size. Finally we determine the largest out-component size. 

The fraction of nodes in the giant out-component $G$ increases monotonically with increasing the node occupancy $p$ as expected (Fig. \ref{fig_p_q}, left panels). For $q=0.3$, $G$ decreases with increasing the network size except for the point at $p=1$. In contrast, for $q=0.6$ and $q=0.7$ there are two regimes. For small $p$ the value of $G$ decreases with increasing $N$, but it remains constant or even increase for $p$ closer to 1. To investigate the size dependence we estimated the slope of the plot $G$ vs $N$ at the different $p$ values (Fig. \ref{fig_p_q}, right panels). For $q=0.3$ the slope ($dG/dN$) is negative for all $p<1$, corroborating that it has a trivial percolation threshold $p_c=1$. In contrast, for $q=0.6$ and 0.7 the slope becomes zero at a value of $p=p_c$, where $0<p_c<1$. The value of $p_c$ is approximately 0.8 for $q=0.6$ and 0.4 for $q=0.7$.

\section{Conclusions}
\label{conclusions}

In conclusion, duplication split-networks are stochastic fractal networks with power-law degree distributions. These networks are fragile for duplication-rates $q<q_c=1/2$: For any node occupation fraction $p<1$ the relative size of the giant component decreases with increasing the network size, i.e. $p_c=1$. That implies that the propagation of risk along these networks can be mitigated by a small number of interventions.

For $q>q_c$ the duplication-split networks have a diverging average out-degree: $\langle k_{out}\rangle\rightarrow\infty$ when $N\rightarrow\infty$. In this regime there is a non-trivial percolation threshold $0<p_c<1$. Based on the numerical simulations for selected values of $q$, we conjecture that $p_c$ changes from $p_c=1$ at $q=1/2$ to $p_c=0$ at $q=1$. The existence of a finite percolation threshold implies that for $q>q_c=1/2$ we need interventions at a finite fraction of nodes to stop the risk propagation.

The behavior for both finite ($q<q_c$) and diverging ($q>q_c$) average out-degree is different from that of random directed networks. For random directed networks there is a percolation threshold at $p_c=1/\langle k_{out}\rangle$, which goes to 0 when $\langle k_{out}\rangle\rightarrow\infty$. The difference is rooted in the fractal vs small-world nature of duplication-split vs random networks, respectively. In fact, the existence of a non-trivial percolation threshold even though $\langle k_{out}\rangle\rightarrow\infty$ was observed previously for a class of deterministic fractal scale-free networks investigated by Rozenfeld and ben-Avraham \cite{rozenfeld07}.

From the point of view of project design, we should avoid activity networks with high effective duplication rate $q$. The effective $q$ can be estimated from the fit of the nodes out-degree to the power law scaling $P(k_{out})\sim k_{out}^{-1/q}$. In the context of construction projects $q$ is concentrated in the range $0.2<q<0.3$, but there are projects with $q$ close and above $q=q_c=1/2$ as well \cite{vazquez03}. For $q>q_c=1/2$ the projects are classified as unmanageable. A delay in one activity can trigger a delay cascade  that cannot be stopped, unless a finite fraction of all downstream activities are mitigated. The latter is either unfeasible of can be achieved at expenses of increasing the project cost dramatically. That is why we call this regimen unmanageable.

Finally, the duplication-split model demonstrates that network growth by duplication can exhibit different structures depending on additional rules. Models of node duplication (or copying) with random pruning or rewiring generate small-world or ultra-small-world networks \cite{kleinberg99, vazquez03dup, chung03, pastor-satorras03,krapivsky05}. In contrast, the split rule tends to increase the paths length between nodes, resulting in fractal networks. When dealing with duplication models, mind the second rule.

\section*{Acknowledgements}

Many thanks to Marcelo Moret and the organizers of CCS2023 Salvador de Bahia, where many of these ideas crystalized. Many thanks to my colleagues Marcio Argollo de Menezes, Javier Borge-Holthoefer and Cristina Masoller for stimulating discussions on the topic. 

This research did not receive any specific grant from funding agencies in the public, commercial, or not-for-profit sectors. Nodes \& Links Ltd provided support in the form of salary for Alexei Vazquez, but did not have any additional role in the conceptualization of the study, analysis, decision to publish, or preparation of the manuscript.

\bibliographystyle{elsarticle-num}

%\bibliography{network.bib}

\end{document}